\documentclass[reprint,amsmath,amssymb,aps]{revtex4-1}
\usepackage[utf8x]{inputenc}
\usepackage{graphicx}
\usepackage{dcolumn}
\usepackage{bm}
\usepackage{color}

\begin{document}

\title{Variational study of the interacting spinless Su-Schrieffer-Heeger model}

\author{M. Yahyavi}
\email{m.yahyavi@bilkent.edu.tr}
\thanks{\\The first two authors contributed equally.}
\affiliation{Department of Physics, Bilkent University, TR-06800
  Bilkent, Ankara, Turkey}
\author{L. Saleem}
\email{luqman.saleem@bilkent.edu.tr}
\thanks{\\The first two authors contributed equally.}
\affiliation{Department of
  Physics, Bilkent University, TR-06800 Bilkent, Ankara, Turkey}
\author{B. Het\'enyi}
\email{hetenyi@fen.bilkent.edu.tr,hetenyi@phy.bme.hu}
\affiliation{Department of Physics, Bilkent University, TR-06800 Bilkent, Ankara, Turkey \\ and \\ MTA-BME Exotic Quantum Phases ``Momentum'' Research Group, Department of Physics, Budapest University of Technology and Economics, H-1111 Budapest, Hungary}

\begin{abstract}
We study the phase diagram and the total polarization distribution of
the Su-Schrieffer-Heeger model with nearest neighbor interaction in
one dimension at half-filling.  To obtain the ground state
wave-function, we extend the Baeriswyl variational wave function to
account for alternating hopping parameters.  The ground state energies
of the variational wave functions compare well to exact
diagonalization results.  For the case of uniform hopping for all
bonds, where it is known that an ideal conductor to insulator
transition takes place at finite interaction, we also find a
transition at an interaction strength somewhat lower than the known
value.  The ideal conductor phase is a Fermi sea.  The phase diagram
in the whole parameter range shows a resemblance to the phase diagram
of the Kane-Mele-Hubbard model.  We also calculate the gauge invariant
cumulants corresponding to the polarization (Zak phase) and use these
to reconstruct the distribution of the polarization.  We calculate the
reconstructed polarization distribution along a path in parameter
space which connects two points with opposite polarization in two
ways.  In one case we cross the metallic phase line, in the other, we
go through only insulating states.  In the former case, the average
polarization changes discontinuously after passing through the
metallic phase line, while in the latter the distribution ``walks
across'' smoothly from one polarization to its opposite.  This state
of affairs suggests that the correlation acts to break the chiral
symmetry of the Su-Schrieffer-Heeger model, in the same way as it
happens when a Rice-Mele onsite potential is turned on.
\end{abstract}

\maketitle

\section{Introduction}

The study of topological systems~\cite{Hasan10,Bernevig13} is an
extremely active research area.  Recent efforts~\cite{Rachel18} have
focused on understanding the effects of electron interaction on such
systems.  Part of this effort concentrates on
extending~\cite{Gurarie11,Wang10,Budich13} the symmetry classification
valid for non-interacting systems~\cite{Altland97,Schnyder08}, part of
it is to map the phase diagrams of existing topological models with
interaction turned on~\cite{Varney10,Jotzu14}.  One model which has
received considerable attention is the Kane-Mele-Hubbard (KMH)
model~\cite{Rachel10} both analytically~\cite{Griset12,Hamad16} and
numerically~\cite{Hohenadler13,Laubach14}.

The starting point in topological analyses is usually the Berry
phase~\cite{Berry84,Shapere89,Xiao10}.  The Berry phase which arises
from integrating across the Brillouin zone~\cite{Zak89} (Zak phase)
corresponds to the polarization~\cite{King-Smith93,Resta94,Resta98} of
a crystalline system, while its modified versions give topological
indices such as the Chern number~\cite{Thouless82} or time-reversal
polarization~\cite{Fu06}.  The Zak phase can be viewed as the first in
a series of gauge invariant
cumulants~\cite{Souza00,Hetenyi14,Yahyavi17}, the second
corresponding~\cite{Resta99} to the variance in the center of mass of
the electronic charge distribution.

Recently, higher order cumulants were
studied~\cite{Patankar18,Kobayashi18,Yahyavi17}.  It was
shown~\cite{Patankar18} that the third cumulant, also known as the
skew, corresponds to the so called shift current, the second-order
{\it nonlinear} optical response in second harmonic generation
experiments (this work addresses such experiments in a Weyl
semimetal).  It was emphasized, that the skew gives a more intuitive
picture of the system, than the sum rules valid for nonlinear
response.  Kobayashi {\it et al.}~\cite{Kobayashi18} study the
quantity $Z^{(q)} = \langle \Psi | \exp(i 2 \pi \hat{X}/L) q | \Psi
\rangle$ ($L$ denotes the system size, $\hat{X}$ is the total position
operator, and $q$ is a real number, integer in systems with periodic
boundary conditions) in the metallic phase.  The $q=1$ case was
suggested by Resta and Sorella~\cite{Resta98,Resta99} for the
polarization and its variance.  A modification to this scheme was
suggested by Aligia and Ortiz~\cite{Aligia99} for lattice systems with
fractional fillings.  It can be shown that for general $q$ this
quantity contains the same information as the gauge invariant
cumulants.  In Ref. \cite{Kobayashi18} it is shown that the size
dependence of $Z^{(q)}$ in the metallic phase is universal.  From the
first six cumulants the distribution of the polarization in the
Rice-Mele (RM) model~\cite{Rice82} was reconstructed~\cite{Yahyavi17}.

In this paper we study the one-dimensional spinless interacting
Su-Schrieffer-Heeger (SSH) model~\cite{Su79}.  The spinful version of
this model was studied by Manmana et al.~\cite{Manmana12}.  So far,
although topological, the spinless version of this model has received
relatively little attention.  We develop a variational approach by
extending the Baeriswyl wave function~\cite{Baeriswyl86,Baeriswyl00}
(BWF) to account for the alternating hoppings of the SSH.  We compare
the ground state energies to exact diagonalization results for small
system sizes, finding excellent agreement.  The phase diagram we find
is remarkably similar to that of the KMH
model~\cite{Rachel18,Rachel10}.  When all hopping parameters are
equal, we find a conducting phase for small interaction (in our
variational treatment a Fermi sea), and a correlated insulator with
charge density wave ordering for large interaction.  When the hoppings
alternate, the small interaction phase is the SSH ground state, but as
the interaction increases, the system tends towards charge density
order, which is weakened by the hopping alternation.  We also
construct a parent Hamiltonian for the BWF type wave function we use.
Our construction allows for plotting the curves traced out by varying
$k$ across the Brillouin zone in the space spanned by the components
of the Hamiltonian.

We then study the behavior of the polarization distribution.  In
particular, we do reconstructions along two paths in the parameter
space of the model, which connect topologically distinct states.  Both
paths connect two states with finite interaction parameter, but with
opposite polarity in hopping.  One path crosses the metallic phase
line (the interaction parameter is constant), while the other passes
through the insulating regime only (the interaction is varied).  We
find that in the first case, the maximum of the distribution remains
constant until the metallic phase line is reached, there the
distribution flattens, and after passing to the other side of the
metallic phase line, the distribution has a maximum at a different
polarization (a jump occurred in the Berry phase at the metallic phase
line).  For the second path, the maximum, as well as the other
cumulants vary smoothly, the distribution ``walks across'' from one
polarization to the other.
\begin{figure}
\includegraphics[width=\linewidth]{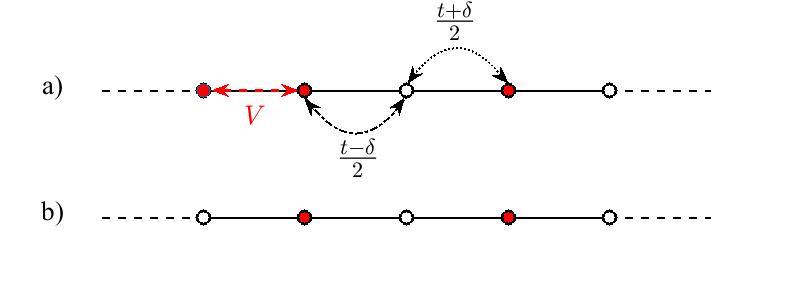}
\caption{(Color online) Graphical representation of our model
    Hamiltonian (part a) ) and the charge density wave (CDW) state
    (part b) ).  Filled(Empty) circles indicate lattice sites
    occupied(not occupied) by particles.  In part a) $V$
    indicates the interaction, $t$ indicates the average hopping,
    $\delta$ denotes the amount by which the hopping alternates
    between even and odd bonds.}
\end{figure}
To make sense of these results, we relate them to the well-known
topological quantum phase transition which occurs in the SSH model.
In the SSH model chiral symmetry gives rise to a symmetry protected
topological phase separated from a topologically trivial phase by gap
closure.  Again we reconstruct the polarization distribution along two
different paths.  Both paths pass between topologically distinct
phases, but along one the symmetries are always respected (gap
closure occurs), while along the other, the symmetries are relaxed.
The polarization distributions evolve exactly in the same manner.

Our paper is organized as follows.  In the following section we give
the models we study, the form of the BWF, the construction of the
parent Hamiltonian, and the gauge invariant cumulants.  In section
\ref{sec:results} our results and analyses are presented.  In section
\ref{sec:conclusion} we conclude our work.

\section{Model and methods}
\subsection{Interacting Su-Schrieffer-Heeger Model}

We study the interacting SSH model in one dimension at half filling.
This model consists of a hopping parameter, which alternates between
odd and even bonds, and an interaction term if two particles are on
nearest neighboring (NN) sites.  For the case of uniform hopping for
all bonds, the model can be solved by the Bethe ansatz, and it is
known that an ideal conductor (finite Drude weight) to insulator phase
transition takes place.

The Hamilton operator of the model consists of two terms, the SSH term
($\hat{h}_{SSH}$) and the interaction term ($\hat{h}_V$) and can
be written as
\begin{eqnarray}
  \label{eqn:Hml}
  \hat{H}=
\overbrace{-\frac{1}{2}\sum_{n=1}^{L} [t+(-1)^n\delta]c_n^\dagger
c_{n+1} + \mbox{H.c.}}^{\hat{h}_{SSH}} \\ \nonumber
+
\overbrace{\frac{V}{2}\sum_{n=1}^{L} n_n n_{n+1}}^{\hat{h}_V}.
\end{eqnarray}
where \(c_n^\dagger\) (\(c_n\)) creates (annihilates) a particle at
site \(n\), \(n_n=c_n^\dagger c_n\) is the density operator at site
\(n\).  \(t\) denotes the average hopping, taken as unity in the
following, \(\delta\) denotes the deviation in hopping between odd
and even sites. \(V\) is the Couloumb interaction between NNs.  For
$\delta = 0$ the ideal conductor insulator transition occurs at
$V=2$.

\subsection{The variational wave function}


The BWF~\cite{Baeriswyl86,Baeriswyl00,Valenzuela03,Hetenyi10} already
has a history of successfully reproducing the properties of strongly
correlated models.  It was originally developed for the fermionic
Hubbard model, but it has been applied to the bosonic
Hubbard~\cite{Hetenyi16} as well as quenches interacting spinless
fermions~\cite{Dora16}.
\begin{figure}
\includegraphics[width=\linewidth]{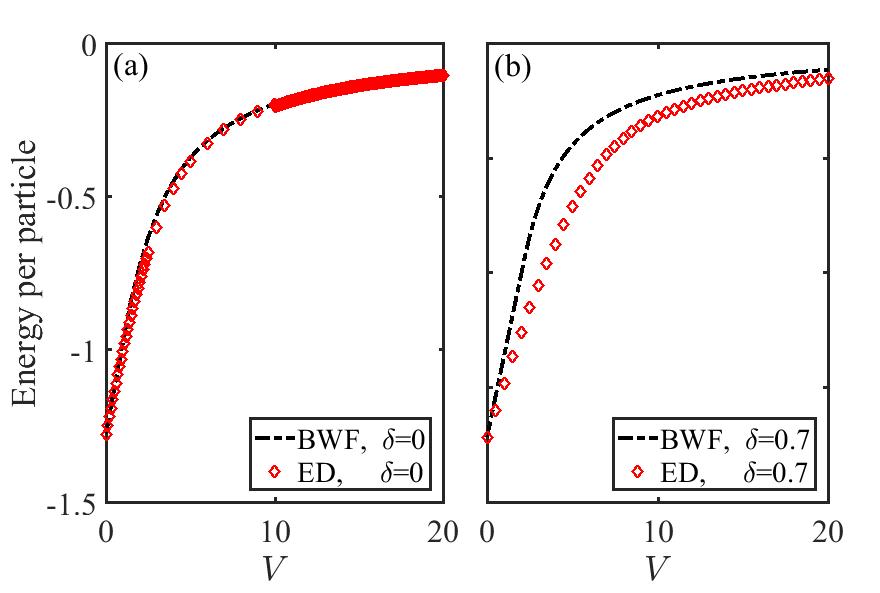}
\includegraphics[width=\linewidth]{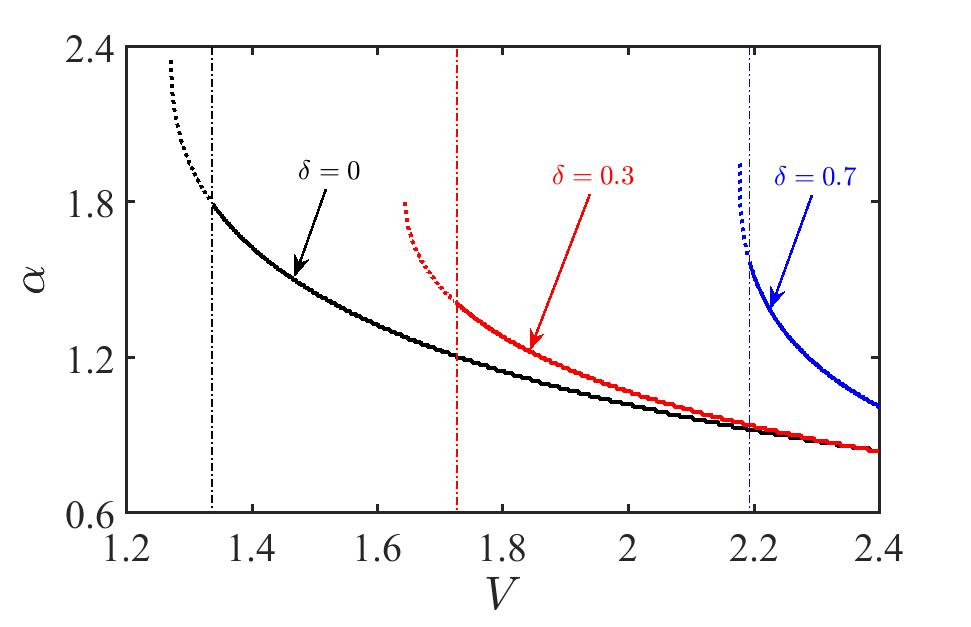}
\caption{(Color online) Upper panel: The variational ground state
  energy per particle for (a) \(\delta = 0,\) and (b) \(\delta= 0.7\)
  based on the Baeriswyl wave function compared to exact
  diagonalization for $12$ lattice sites .  Lower panel: the
  variational parameter as a function of interaction strength for
  \(\delta = 0, \pm 0.3\) and \(\pm 0.7\).  Solid lines indicate the
  global minimum in the CDW type insulating phase.  Dashed lines
  indicate metastable insulating phases on the metallic side of the
  phase diagram.}
	\label{fig:BWFvsED}
\end{figure}

In this section we describe the BWF~\cite{Baeriswyl86,Baeriswyl00},
its extensions necessary to account for the alternating hoppings of
the SSH model, and write down the solution for the variational energy
in closed form.
\begin{figure}
		\includegraphics[width=\linewidth]{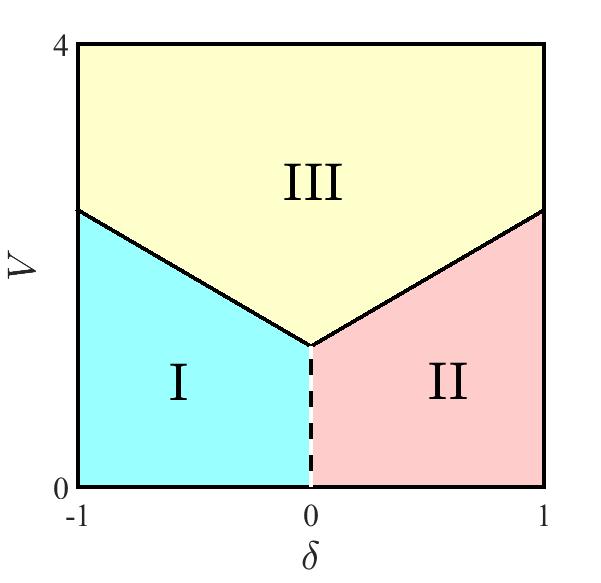}
	\caption{(Color online) Phase diagram.  $I$ and $II$ are SSH
          states (Hartree-Fock approximation).  $III$ is a CDW type
          phase, with finite skew when $\delta$ differs from zero.
          The dashed line indicates a Fermi sea (ideal conductor).}
	\label{fig:PD}
\end{figure}

The BWF starts with the wave function at infinite interaction (a
charge-density wave) and acts on this wave function with a kinetic
energy based projector.  In analogy to this idea we write
\begin{equation}
  \label{eqn:BWF1}
|\Psi_B (\alpha)\rangle = N_B \exp(-\alpha \hat{h}_{SSH})|\Psi_\infty\rangle,
\end{equation}
where $\alpha$ denotes the variational parameter and
$|\Psi_\infty\rangle$ stands for a perfectly ordered charge density
wave, and $N_B$ is a normalization constant.  Our further calculations
are rendered easier by the particularly simple form of $|\Psi_\infty\rangle$,
\begin{equation}
  \label{eqn:Psi_inf}
|\Psi_\infty\rangle=\underset{k\epsilon RBZ}{\prod}
        \frac{1}{\sqrt{2}}(c_k^\dagger + c_{k+Q}^\dagger )|0\rangle
\end{equation}
where $Q=\pi$ is ordering wave vector (the lattice constant was
assumed to be unity), the product runs through the reduced Brillouin
zone (RBZ) and $|0\rangle$ is fermionic vacuum state.  Fourier
transforming the Hamiltonian results in,
\begin{eqnarray}
\label{eq:H_k}
		\hat{h}_{SSH}(k) & =
		\begin{bmatrix}
			c_k^\dagger &  & c_{k+\pi}^\dagger
		\end{bmatrix}
		\begin{bmatrix}
			\epsilon(k) &  & i\gamma(k) \\-i\gamma(k)&&-\epsilon(k)
		\end{bmatrix}
		\begin{bmatrix}
			c_k \\c_{k+\pi}
		\end{bmatrix}                                                                                     \\
		\hat{h}_V & =-\frac{V}{L}\sum_{k,k',q} \epsilon(q)  c_{k+q}^\dagger c_k c_{k'-q}^\dagger c_{k'}
\end{eqnarray}
where $\epsilon(k)=-t\cos{k},\gamma(k) =-\delta \sin{k}$.  With the
help of the Pauli matrix representation of the SSH noninteracting
Hamiltonian,
\begin{equation}
  \hat{h}_{SSH}(k) = \epsilon(k) \sigma_z - \gamma(k) \sigma_y
\end{equation}
the SSH Hamiltonian based projector can be expanded as
\begin{equation}
 \label{eqn:h_ssh_exp}
\exp(-\alpha \hat{h}_{SSH}(k))=\cosh(\alpha
h(k))\hat{\hat{I}}-\sinh(\alpha h(k))\hat{\hat{h}}_{SSH}(k)
\end{equation}
where $h(k)=\sqrt{t^2\cos^2k+\delta^2 \sin^2 k}$,
$\hat{\hat{I}}$ is the 2-by-2 identity matrix, and
\begin{equation}
  \hat{\hat{h}}_{SSH}(k) = \frac{\hat{h}_{SSH}(k)}{h(k)},
\end{equation}
a unit vector in the space spanned by $\epsilon(k)$ and $\gamma(k)$.
Now, applying the projector \eqref{eqn:h_ssh_exp} on the CDW state (Eq. 
\eqref{eqn:Psi_inf}) we obtain the normalized wave function,
\begin{equation}
  \label{eqn:Psi_nrm}
|\Psi_B(\alpha)\rangle=\underset{k\epsilon RBZ}{\prod} \frac{A(k) c_k^\dagger + A(k+Q) c_{k+Q}^\dagger}{\sqrt{2\cosh{[2\alpha h(k)]}}}|0\rangle,
\end{equation}
with
\begin{equation}
  \label{eqn:A_k}
	A(k) =\cosh{[\alpha h(k)]}-\frac{\sinh{[\alpha h(k)]}}{h(k)}\epsilon(k)+i\gamma(k)\frac{\sinh{[\alpha h(k)]}}{h(k)}.
\end{equation}
\begin{figure}
\includegraphics[width=\linewidth]{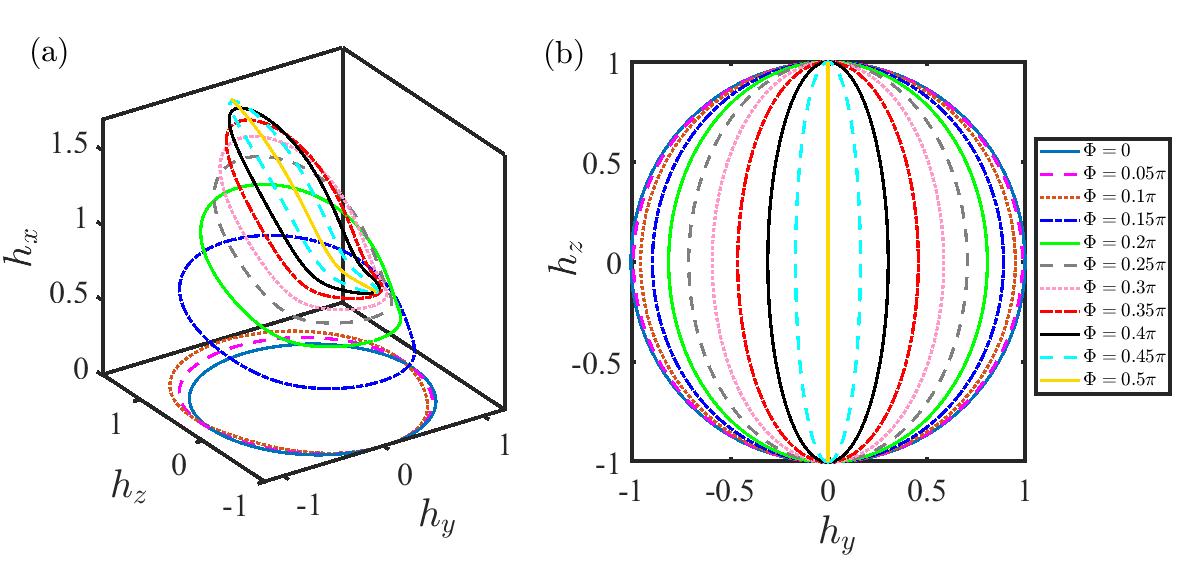}
\caption{(Color online) Curves traced out by the Hamiltonian
          in the Brillouin zone.  Left panel shows the correlated
          system, $\delta = \cos(\phi), V = 1 + 4\sin(\phi)$.  On the
          right panel $\delta = \cos(\phi), V = 1$.  For both panels,
          the values of the variable $\phi$ are indicated in the
          legend on the right.  }
	\label{fig:BZcurves}
\end{figure}

Having derived the action of the projector on the CDW state, we can
now evaluate the variational estimate for the ground state energy by
calculating the expectation value of Eq. (\ref{eq:H_k}) over our
extended Baeriswyl wave function.  We give the expectation values of the SSH Hamiltonian and of the interaction separately as
\begin{equation}
  \label{eqn:h_SSH}
  \langle \hat{h}_{SSH} \rangle = -\sum_k h(k)\tanh{[2\alpha h(k)]}
\end{equation}
and
\begin{equation}
  \label{eqn:h_V}
  \langle \hat{h}_{V} \rangle = \frac{VL}{4} -\frac{V}{L}\sum_{i=1}^3|T_i|^2
\end{equation}
where
\begin{eqnarray}
  \label{eqn:h_V_sums}
T_1 = & \sum_k \bigg[ \frac{1}{2\cosh{[2\alpha h(k)]}}
  +\frac{i\gamma(k)}{2h(k)} \tanh{[2\alpha h(k)]} \bigg] \\ \nonumber
T_2 = & \sum_{k}\bigg[ \frac{\epsilon(k)}{2} -\frac{\epsilon(k)^2}{2h(k)}
  \tanh{[2\alpha h(k)]} \bigg] \\ \nonumber
T_3 = & \sum_{k} \bigg[ \frac{\epsilon(k)}{2\cosh{[2\alpha h(k)]}}
  +\frac{i\epsilon(k)\gamma(k)}{2h(k)} \tanh{[2\alpha h(k)]} \bigg].
\end{eqnarray}
For the $t-V$ model these expressions were derived in Ref. \cite{Dora16}.

\begin{figure}
\includegraphics[width=\linewidth]{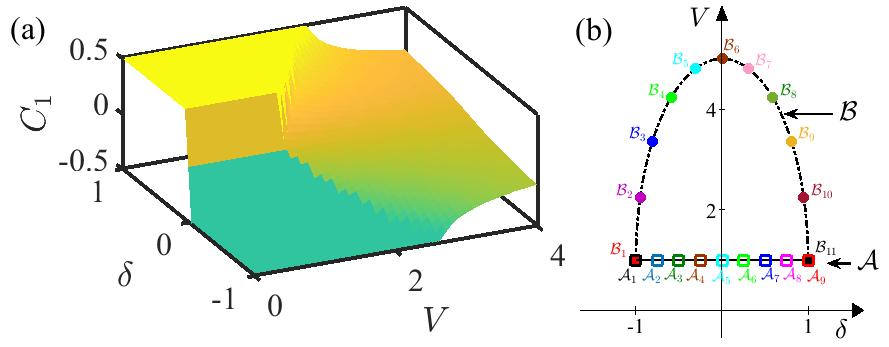}
\includegraphics[width=\linewidth]{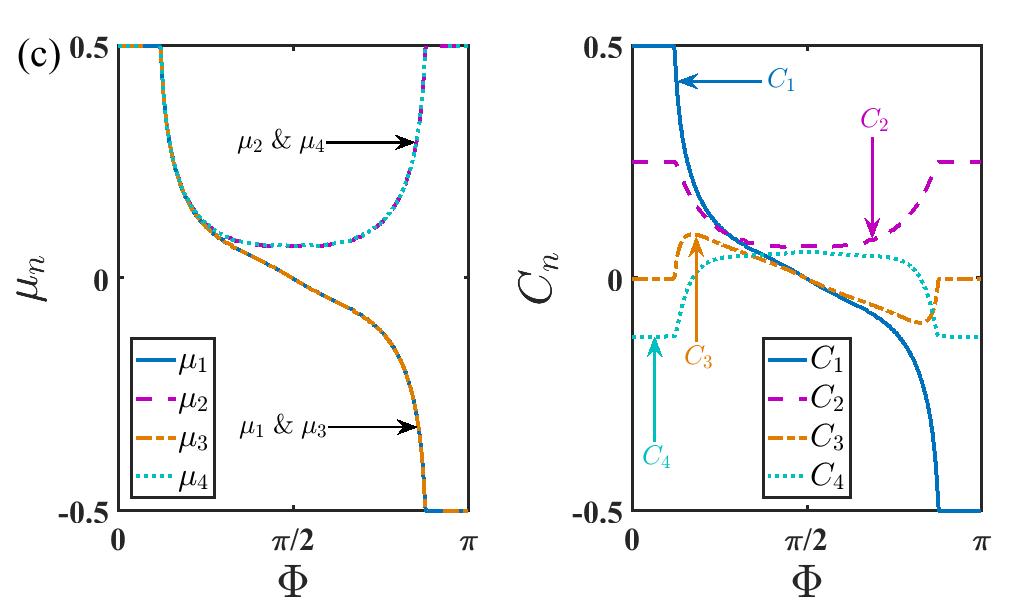}
\includegraphics[width=\linewidth]{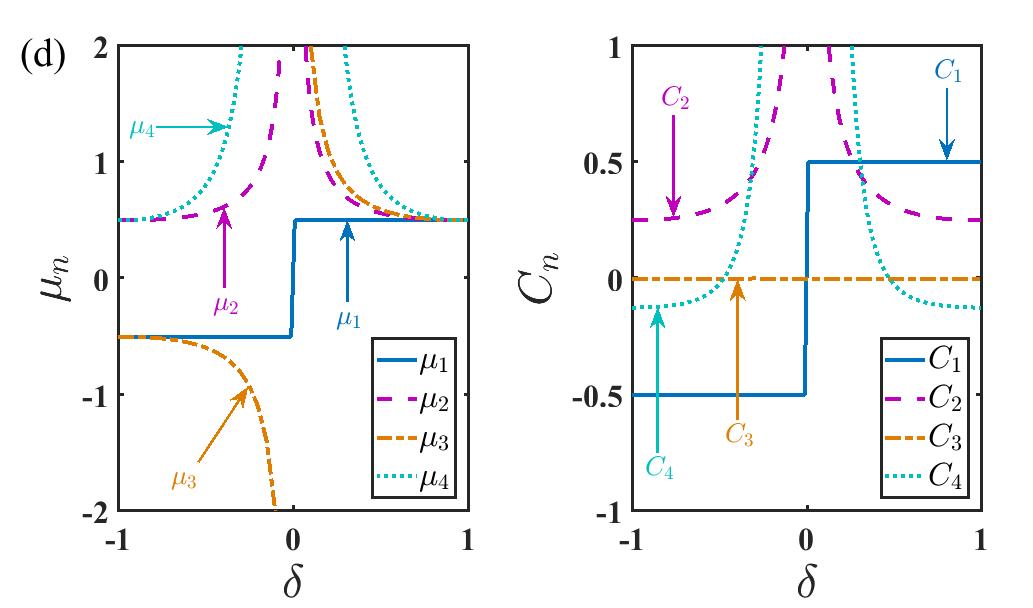}
	\caption{(Color online) (a) First cumulant, ($C_1$) or
          polarization, of the interacing Su-Schrieffer-Heeger model
          as a function of $\delta$ and $V$.  (b) Lines in the
          $\delta-V$ plane for which we calculate the cumulants and
          reconstruct the polarization.  Two paths are shown between
          the points $-1,1$ and $1,1$: path $\mathcal{A}$ is a
          straight line, path $\mathcal{B}$ is a semi-ellipse.  (c)
          Moments and cumulants along the semi-elliptic
          ($\mathcal{B}$) path.  (d) Moments and cumulants along the
          straight line ($\mathcal{A}$) path. }
	\label{fig:Cmlnts_ellps_ln}
\end{figure}

This newly derived wave function gives exact results in two extreme
limits, $V\rightarrow 0$ and $V\rightarrow \infty$. In the former
the total energy becomes equal to the energy of the SSH model, while
in the latter the total energy tends to zero. In between these two
opposite extreme limits a comparison of numerically calculated total
energy with exact diagonalized calculation and optimal minimization
parameter is given in the upper panel of Fig. \ref{fig:BWFvsED} for a
small sized system.  The extended BWF gives results which compare well
in all cases, although as $\delta$ increases, the agreement worsens.

\subsection{Parent Hamiltonian of the Baeriswyl wave function}

\label{sec:parH}

Visualizing topological phase transitions is greatly aided in the case
of Hamiltonians which are two-state in $k$-space via plotting the
curves traced out by sweeping through the Brillouin
zone~\cite{Asboth16}.  The standard way is to write the two-state
Hamiltonian in the form $H(k) = {\bf h}(k)\cdot \sigma$, where ${\bf
  h}(k)$ is a three-dimensional vector which traces out a closed curve
in the space spanned by $h_x(k)$, $h_y(k)$, $h_z(k)$.

We can use the results of the previous subsection to construct an
effective parent Hamiltonian for the BWF, which is also of the form
$H(k) = {\bf h}(k)\cdot \sigma$.  Our construction consists of two
steps.  First, inspecting Eqs. (\ref{eqn:h_SSH})-(\ref{eqn:h_V_sums}),
we are able to write the total energy (apart from the constant shift
$VL/4$) as a {\it single} sum over $k$, as $E = \sum_k
\tilde{\epsilon}_k$.  Since Eqs. (\ref{eqn:h_V_sums}) include double
$k$-sums, these contributions to $\tilde{\epsilon}(k)$ appear as
single $k$-sums.  Second, we consider the coefficients at $k$ and
$k+Q$, and obtain the angles $\theta_k$ and $\phi_k$ according to the
normalized equation 
\begin{eqnarray}
   \nonumber
\tilde{\epsilon}_k \begin{bmatrix}
			\cos(\theta_k) & & \sin(\theta_k)e^{-i \phi_k} \\
                        \sin(\theta_k)e^{i \phi_k} & &-\cos(\theta_k)
		\end{bmatrix}
		\begin{bmatrix}
			\frac{\sin(\theta_k)}{2}  \\
                        -\frac{\cos(\theta_k)}{2} e^{i \phi_k}
		\end{bmatrix}
                = \\ - \tilde{\epsilon}_k
		\begin{bmatrix}
			\frac{\sin(\theta_k)}{2}  \\
                        -\frac{\cos(\theta_k)}{2} e^{i \phi_k}
		\end{bmatrix},
\end{eqnarray}
where we equate 
\begin{eqnarray}
  \frac{A(k)}{\sqrt{2\cosh{[2\alpha h(k)]}}} &=& \frac{\sin(\theta_k)}{2} \\
  \frac{A(k+Q)}{\sqrt{2\cosh{[2\alpha h(k)]}}} &=& -\frac{\cos(\theta_k)}{2} e^{i \phi_k} . \nonumber
\end{eqnarray}
This procedure guarantees that summing over the RBZ gives the correct
ground state energy, and that at each $k$-vector, the correct
coefficients $A(k)$ and $A(k+Q)$ are obtained.

\subsection{Polarization and gauge invariant cumulants}

The gauge invariant cumulant series associated with the polarization
was first studied by Souza, Wilkens, Martin~\cite{Souza00}.  In a
general sense it can be derived~\cite{Hetenyi14} based on the discrete
Berry phase (Bargmann invariant~\cite{Bargmann64}).  Here we give the
basic expressions for the gauge invariant cumulants, for the
reconstruction of the polarization we refer the reader to
Ref. \cite{Yahyavi17}.

\begin{figure}
\includegraphics[width=\linewidth]{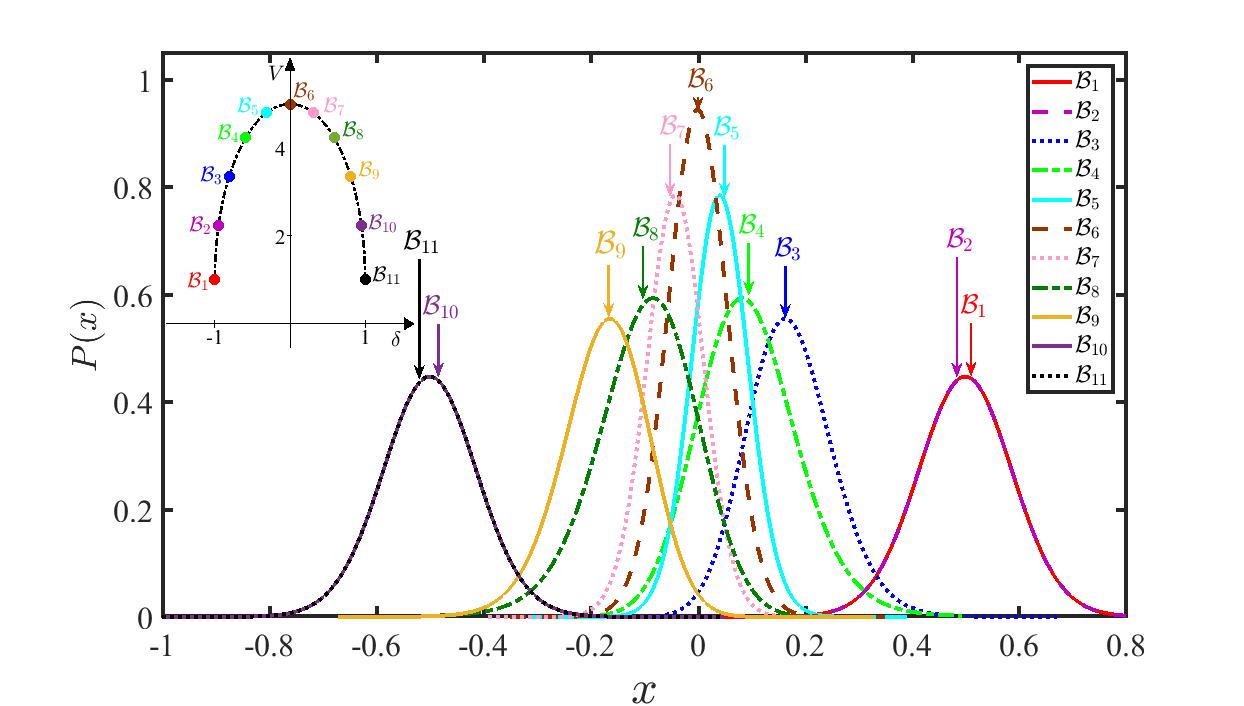}
\includegraphics[width=\linewidth]{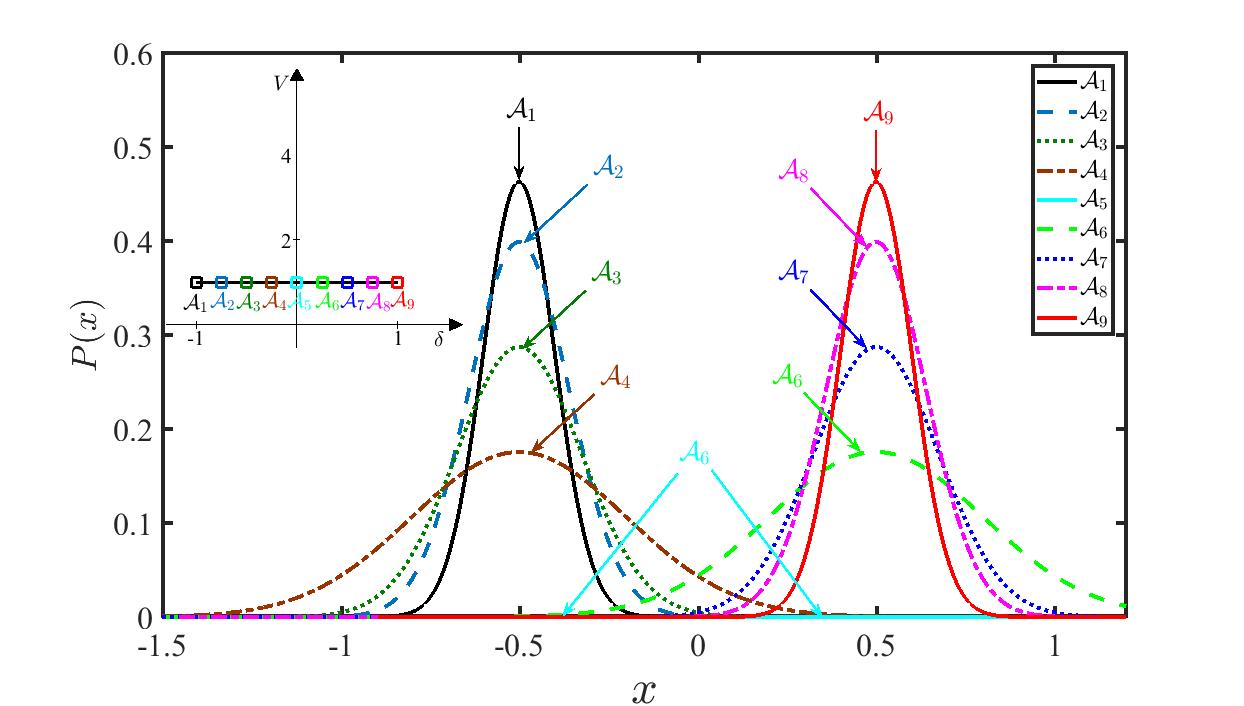}
\caption{(Color online) Reconstructed polarization distributions along
  the two paths shown indicated in Fig. \ref{fig:Cmlnts_ellps_ln}.
  The upper panel(lower panel) shows reconstructed polarizations along
  the ellipse(straight line) between the points $-1,1$ to $1,1$ on the
  $\delta-V$ plane.}
\label{fig:Pol_VSSH}
\end{figure}

Consider a one-dimensional system whose Hamiltonian is periodic
in $L$.  Taking periodic Bloch functions parametrized by the crystal
momentum, $u(k)$, and defining $\gamma_m(k) = \langle u(k) |
\partial^m_k | u(k)\rangle$, the first four cumulants take the form
\begin{eqnarray}
\label{eqn:cmlnts}
C_1 &=& i \frac{L}{2\pi} \int_{-\frac{\pi}{L}}^{\frac{\pi}{L}} d K
\gamma_1 \\ \nonumber C_2 &=& -\frac{L}{2\pi}
\int_{-\frac{\pi}{L}}^{\frac{\pi}{L}} d K [\gamma_2 - \gamma_1^2]
\\ \nonumber C_3 &=& -i \frac{L}{2\pi}
\int_{-\frac{\pi}{L}}^{\frac{\pi}{L}} d K [\gamma_3 -3 \gamma_2
  \gamma_1+ 2\gamma_1^3] \\ \nonumber C_4 &=& \frac{L}{2\pi}
\int_{-\frac{\pi}{L}}^{\frac{\pi}{L}} d K [\gamma_4 -3 \gamma_2^2
  -4\gamma_3\gamma_1 + 12 \gamma_1^2\gamma_2 -6\gamma_1^4].
\end{eqnarray}
These quantities can be shown to be gauge invariant.  If the Wannier
functions associated with the Bloch functions are sufficiently
localized, they correspond to the cumumlants of the probability
distribution of the total position, and can be used in its
reconstruction~\cite{Yahyavi17}.  From the inversion of this cumulant
series it is also possible to obtain gauge invariant moments.  The
cumulants can be inverted to obtain the gauge invariant moments,
\begin{eqnarray}
\label{eqn:muC}
\mu_C^{(1)} &=& C_1 \\ \nonumber \mu_C^{(2)} &=& C_2 + C_1^2
\\ \nonumber \mu_C^{(3)} &=& C_3 + 3 C_2 C_1 + C_1^3 \\ \nonumber
\mu_C^{(4)} &=& C_4 + 4 C_3 C_1 + 3 C_2^2 + 6 C_2 C_1^2 + C_1^4.
\end{eqnarray}

In our model the wave function is of the form given in
Eq. (\ref{eqn:Psi_nrm}).  The cumulants can be obtained in a
straightforward way, for example,
\begin{equation}
\gamma_m(k) =  A^*(k) \partial_k^m A(k) + A^*(k+Q) \partial_k^m A(k+Q).
\end{equation}
Using $\gamma_m(k)$ the cumulants can be constructed according to
Eq. (\ref{eqn:cmlnts}), however the integrals and the normalization
are now over the RBZ.
\begin{figure}
\includegraphics[width=\linewidth]{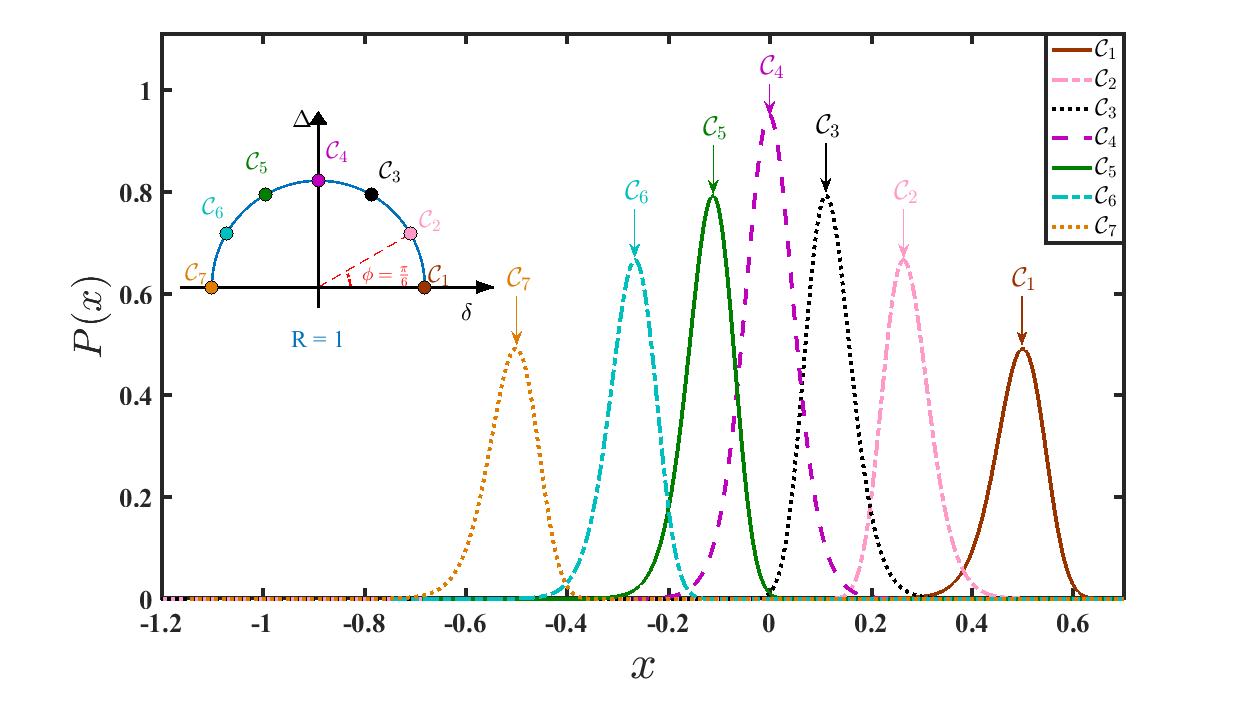}
\includegraphics[width=\linewidth]{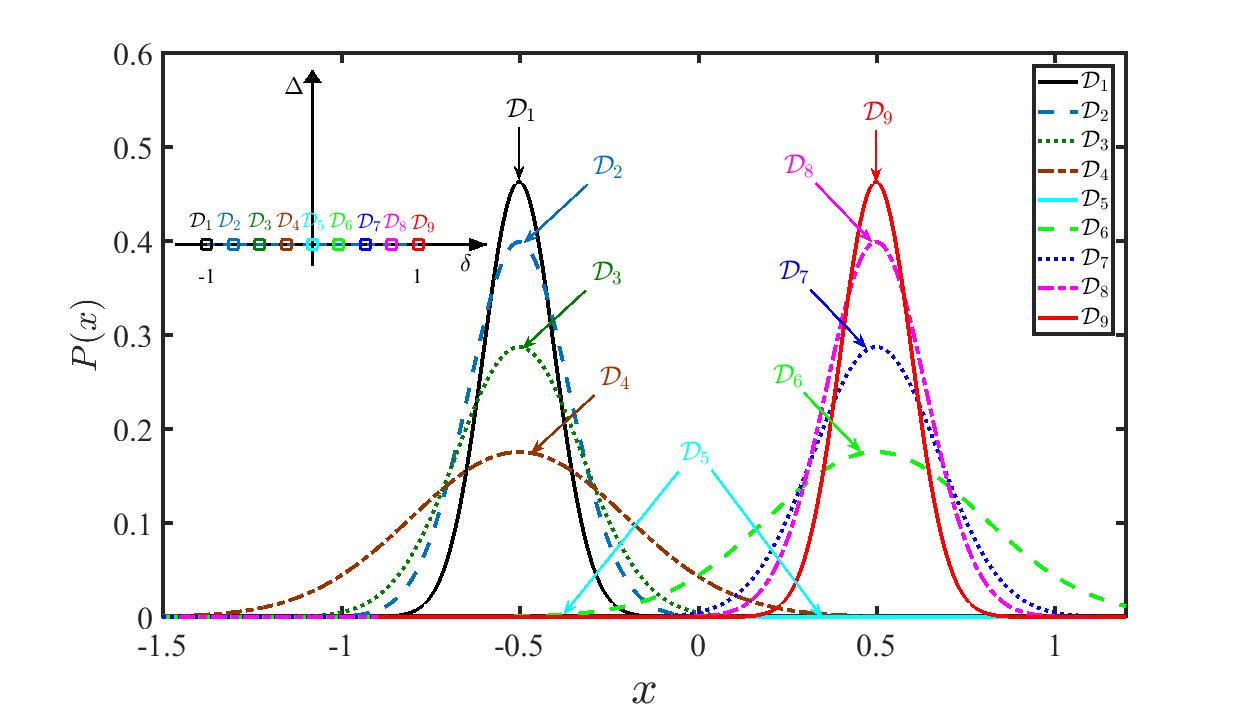}
\caption{(Color online) Reconstructed polarization for the Rice-Mele
  model along the paths indicated in the inset of each plot.  The
  paths are represented on the $\delta-\Delta$ plane.}
\label{fig:RM}
\end{figure}

\section{Results and analysis}

\label{sec:results}

The phase diagram of the model according to our calculations is shown
in Fig. \ref{fig:PD}.  For all values of $\delta$ the variational
parameter at which the total energy is minimized is finite for large
$V$ indicating a CDW type insulating phase.  At some $V_t$ for each
$\delta$, the minimum at finite $V$ becomes a local minimum, with
$\alpha=\infty$ becoming the global minimum, meaning that within our
variational approximation the phase transition is first order (see
Fig. \ref{fig:BWFvsED}).  The small $V$ phase at $\delta=0$ is a Fermi
sea, while for finite $\delta$ it is an SSH state.  The metastable
phase disappears at some finite $\delta$.  Phase $III$ is an
insulating phase.  When $\delta=0$ the skew is zero in this phase,
when it is finite, a finite skew develops (see
Fig. \ref{fig:Cmlnts_ellps_ln}).  Our phase diagram is qualitatively
similar to the two-dimensional KMH model, where the topologically
non-trivial phases at small interaction become magnetically ordered
insulating phases~\cite{Rachel18,Rachel10}.

We also calculated the curves traced out by the vector ${\bf d}$ which
defines the parent Hamiltonian derived in section \ref{sec:parH},
shown in Fig. \ref{fig:BZcurves}.  The left panel (part (a)) shows the
curve traced out by the Brillouin zone along specific points of a
curve in the parameter space of the Hamiltonian which starts at
$\delta=-1,V=1$ and ends at $\delta=1$ and $V=1$.  The curve in
between is an ellipse which does not cross the metallic phase line.
The curves traced out our cyclic in three dimensions.  At $\delta=0$
is approached the cyclic curves become ``thinner'' and at $\delta=0$
itself the BZ is represented by a line rather than a cyclic curve.
The right panel (part (b)) shows what happens along the straight line
$V=1$ but $\delta$ varying from minus one to one.  This is exactly
what happens in the SSH model, whose Hamiltonian is $d_z = -t \cos(k)$
and $d_y = -\delta sin(k)$.  As $\delta$ approaches to zero (gap
closure) the curve becomes a line along the $y$-axis.  The left panel
gives some indication of the effect of correlation.  Standard
mean-field theory of the SSH model gives the RM model corresponding to
an additional term in the $d_x$ directions with a renormalized on-site
potential strength, and whose curves (not shown) would be similar to
panel (b) of Fig. \ref{fig:BZcurves}, but with a shift in the $d_x$
direction.  One difference between mean-field theory and our BWF based
treatment can be seen in the curves of panel (a) of
Fig. \ref{fig:BZcurves}.

In part (a) of Fig. \ref{fig:Cmlnts_ellps_ln} we show the polarization
($C_1$) on the $\delta-V$ plane.  The polarization is take to be zero
in the limit $V \rightarrow \infty$.  As $V$ decreases the absolute
value of the polarization increases, its sign depends on the sign of
$\delta$.  For finite $\delta$ the polarization saturates at a finite
value of $\delta$.  Particularly interesting is the behavior at
$\delta=0$, where the polarization remains the same value, but we see
that below $V<1.3365...$ it rises rapidly from $\delta<0$ to $\delta>0$,
almost discontinuously.  For $V>1.3365...$, the rise in $C_1$ across
$\delta=0$ is smooth.

In part (b) of this Figure we show two paths we have chosen for our
subsequent analysis.  Paths $\mathcal{A}$ and $\mathcal{B}$ both
connect the points $-1,1$ and $1,1$ on the $\delta-V$ plane, but path
$\mathcal{A}$ crosses the line segment $\delta=0$, $0<V<1.3365...$,
while $\mathcal{B}$ does not, it passes $\delta=0$ above the point
$V=1.3365...$.  For the elliptical path the moments and cumulants are
shown in part (c) of the Figure.  All cumulants and moments change
continuously $C_2$ is minimum at $\delta=0$.  In contrast to this,
along the linear path $C_1$ changes discontinuously.  $C_2$ increases
sharply around $\delta=0$, indicating delocalization.  All the
cumulants change rapidly around $\delta=0$, the even cumulants are
even functions, while the odd ones are odd.

In Fig. \ref{fig:Pol_VSSH} we show reconstructed polarization
distributions along a chosen set of points along the two different
paths $\mathcal{A}$ and $\mathcal{B}$.  Along the elliptical path
($\mathcal{B}$) the polarization distribution ``walks across''
smoothly between the two positions related by symmetry.  The
distribution is a smooth function, the maximum changes continuously,
and the shape of the distribution indicates a localized state
(insulating).  In contrast to this, when the polarization is
reconstructed along the line which crosses the line
$\delta=0,0<V<1.3365...$, the maximum of the distribution remains fixed in
the interval $\delta>0$, while the width of the distribution is
increasing (the distribution is becoming more delocalized).  At
$\delta=0$ the distribution is flat, the system is delocalized.  We
interpret this as a conducting state.  For $\delta<0$ the distribution
localizes around a different maximum.  As the absolute value of
$\delta$ increases the distribution becomes more localized.

We can relate the effect of the correlation to what happens around the
topologically nontrivial point~\cite{Xiao10} of the SSH and RM models.
We take as the definition of the RM model to be the SSH model defined
above, plus an alternating on-site potential, which breaks the chiral
symmetry, of the form $\Delta \sum_j (-1)^j c_J^\dagger c_j$, where
$\Delta$ denotes the strength of the potential.  In Fig. \ref{fig:RM}
we show reconstructed polarizations along a semi-circle on the
$\delta-\Delta$ plane, and along a line along which the model is SSH,
with a topological phase transition protected by symmetry.  The gap
closure (phase transition) point is at $\delta=\Delta=0$.  The
symmetry broken semi-circular path shows an evolution of the
polarization distribution similar to the upper panel in
Fig. \ref{fig:Pol_VSSH} with the maximum shifting continuously, while
along the linear path, the maximum shifts discontinuously between the
$\delta<0$ and $\delta>0$ cases.

\section{Conclusion}

\label{sec:conclusion}

We have developed a variational theory based on the Baeriswyl wave
function for the SSH model with interaction, calculated the phase
diagram f the model, and have studied the behavior of the polarization
distribution as a function of the interaction.  The phase diagram of
the variational approach is particularly interesting: for the
interacting system with homogeneous hopping a conductor-insulator
transition is produced at $V=1.3365...$, rather than $V=2$ (which is
the result of the exact solution).  In our approximate scheme the
transition is first order.  Overall, the phase diagram is similar to
the one found for another interacting topologial model, the
Kane-Mele-Hubbard model.~\cite{Rachel18,Rachel10} The Gutzwiller wave
function in this case can only produce a metallic
state~\cite{Millis91}.

Our study of the reconstructed polarization shows the effect of
correlation on the phase diagram of the Su-Schrieffer-Heeger model.
It is instructive to compare the situation to the Rice-Mele model,
which is an SSH model with an on-site potential.  This model exhibits
a topologically non-trivial point (gap closure) at $\delta=0$
($\delta$ being the alternation in hoppings between odd and even
bonds) and zero onsite potential.  When interaction is added to the
SSH model, the gap closure region is extended, it becomes a line,
rather than just a point.  Strictly speaking, gap closure is not
accessible in our variational formalism, but the behaviour of the
reconstructed polarization is identical.  When a gap closure point, or
line, is crossed, the polarization changes discontinuously, while if
two states are connected by a path at which the gap does not close,
the polarization evolves continuously.  While our approach is
variational, therefore approximate, we expect the qualitative picture
to be robust, since the interaction gives rise to a symmetry broken
state which is similar to the ground state of the Rice-Mele model.  In
the Rice-Mele model the charge density wave results from the
alternating on-site potential.  While, broadly speaking, our results
suggests that turning on the interaction in the SSH model gives rise
to similar effects as applying an alternating on-site interaction
(Rice-Mele), the difference is that the gap closure point of the
non-interacting model becomes a line of points in the presence of
interaction (also known from the exact solution of the model), while
it is still a point in the parameter space in the RM case.

\section*{Acknowledgments}
This research was supported by the National Research, Development and
Innovation Fund of Hungary within the Quantum Technology National
Excellence Program (Project Nr.  2017-1.2.1-NKP-2017-00001).  BH
thanks Bal\'azs D\'ora for helpful discussions.

\end{document}